# Evaluating Uncertainties in CFD Simulations of Patient-Specific Aorta Models using Grid Convergence Index Method

O. Aycan[a,b,∗], A. Topuz[a], L. Kadem[b]

[a]Dept. of Mechanical Engineering, Faculty of Engineering, Zonguldak Bulent Ecevit University, Zonguldak, Turkey
[b]Laboratory of Cardiovascular Fluid Dynamics, Department of Mechanical Industrial and Aerospace Engineering, Concordia University, Montreal, QC, Canada

**Abstract**

Cardiovascular diseases are among the most important causes of global mortality. Computational Fluid Dynamics (CFD) is a powerful research tool that analyzes the hemodynamics of artery and blood flow patterns. In this study, CFD simulations are performed to assess the patient-specific healthy aorta, fusiform, and saccular aneurysm with various mesh types, including tetrahedral, polyhedral, and poly-hexacore. The aim of this study is to explore how different mesh types and grid densities impact the hemodynamic properties of physiological flows, with the goal of identifying the most cost-effective meshing approach. A mesh independence study is carried out to ensure the precision of the results, considering the wall shear stress distribution. For this, five different mesh resolutions are generated for each geometry. The uncertainties of the simulations associated with the discretization techniques and solutions are evaluated using the Grid Convergence Index (GCI) method. The findings showed that increasing the mesh density provides smaller uncertainty. GCI values for the wall shear stress are in the range of convergence, indicating that the results are reliable and accurate. Mesh type selection affects the accuracy and computational cost of our simulations. The polyhedral and poly-hexacore meshes lead to a good compromise between precision and computational cost, while the tetrahedral mesh style gives the most precise results with fluctuation. This work provides a systematic approach based on the Grid Convergence Index method in order to select the most appropriate mesh type for evaluating uncertainties in CFD simulations of patient-specific healthy aortas and aortas with abdominal aneurysms. According to the findings and GCI analysis, the polyhedral mesh type was chosen for all patient-specific aorta models. The study clearly demonstrated its superiority over other mesh types, considering uncertainties and computational costs associated with different mesh styles.

## 1. Introduction

Evaluating the hemodynamic effects that may trigger the formation of Abdominal Aortic Aneurysms (AAAs), a relatively common cardiovascular disease that can lead to rupture, is crucial and requires an interdisciplinary approach. Studies have shown that shear stress on the arterial wall can cause bleeding and rupture of AAAs [1]. In addition to wall shear stresses, the blood flow rate is also a factor that can contribute to arteriosclerosis [2]. Low shear stresses and their impact on the arterial wall are associated with the formation of thrombi (intravascular clots) and intravascular platelets [3–5]. Therefore, characterizing the flow characteristics in the abdominal aorta provides critical insights into the risk of AAA formation, development, and rupture.

Experimental approaches utilize physical models and advanced instrumentation to measure and observe blood flow behavior inside the artery, and they are typically more accurate than computational fluid dynamics (CFD) methods. However, CFD can simulate complex three-dimensional flow phenomena and conditions that are challenging to capture in laboratory settings. Furthermore, CFD simulations are faster and more cost-effective than experimental studies. Many studies have employed various numerical methods to examine flow in aortic models and investigate flow characteristics that may trigger the formation or progression of AAA [6, 7].

Different CFD methods provide detailed information about flow structure, pressure values, wall characteristics, and viscous shear stresses in the abdominal aorta. However, selecting the appropriate mesh type and size is a major concern due to its significant impact on the results, as highlighted in previous studies [8, 9]. Achieving mesh independence for the velocity field, using both adapted and non-adapted meshes, is relatively straightforward. However, this is not the case for the evaluation of wall shear stress [10]. Consequently, objectively selecting the most suitable mesh type and size remains one of the most challenging tasks when performing patient-specific CFD simulations.

∗Corresponding author. Tel.: +90-372-291-1144.
*Email address:* osman.aycan@beun.edu.tr; osman.aycan@concordia.ca (O. Aycan)

To obtain reliable results in CFD, it is critical to thoroughly evaluate and minimize spatial discretization errors arising from the resolution of the spatial grid [11, 12]. To address this issue, Roache [13] has proposed the use of the Grid Convergence Index (GCI) Method to facilitate consistent reporting of grid refinement studies in Computational Fluid Dynamics. This technique provides an objective and asymptotic approach to quantify the uncertainty associated with grid convergence.

In previous studies [14–16], CFD uncertainties and discretization errors of a curved tube representing aortic stenosis and coarctation of the aorta and a patient-specific ascending thoracic aortic aneurysm (ATAA), as well as an aorta model with a bicuspid aortic valve, were evaluated using the GCI method. The evaluation focused on two criteria: wall shear stress and velocity. The geometries were discretized using hexahedral and unstructured tetrahedral element types. The GCI method was also applied to bifurcation airway models to assess the performance of grid convergence and different discretization techniques, including structured and unstructured hexahedral, prismatic, unstructured, and adaptive tetrahedral, and hybrid grids, based on velocity profiles and particle deposition fractions. Some uncertainty has been reported regarding the variety of mesh sizes, indicating that the mesh is not in the asymptotic region [17, 18].

Using similar double bifurcation respiratory models and a 90°-bend model with extensions, researchers investigated the effects of a relatively new polyhedral element type compared to 4- and 5-block hexahedral meshes in predicting aerosol deposition. Although the solution convergence time with polyhedral elements was 50% to 140% longer compared to hexahedral meshes of similar size, the polyhedral mesh style showed promise as an excellent alternative to the widely accepted hexahedral mesh style, providing adequate resolution, especially near walls [19]. Lotfi et al. [9] also evaluated different mesh configurations, considering idealized and stented artery models have been evaluated, and the uncertainties of different meshing styles have been calculated based on the wall shear stress near the stent wall using the grid convergence index method. The flow-adaptive polyhedral mesh has shown better results compared to conventional adaptive and nonadaptive tetrahedral meshes.

As a result, powerful computational infrastructures are necessary to accurately solve complex geometries, such as patient-specific aorta models. This issue remains a significant constraint in conducting biofluid simulations and translating CFD simulations into clinical practice. For such complex configurations, selecting the mesh that achieves the best balance between accuracy and computational time becomes crucial. To the best of the authors' knowledge, no study has compared different discretization techniques and their performance in patient-specific geometries based on the GCI method.

This study aims to apply three different discretization techniques to patient-specific aorta models: a healthy aorta, an aorta with a fusiform aneurysm, and an aorta with a saccular aneurysm. The computational performance of these methods will be assessed based on the wall shear stress by conducting a mesh independence study with different cell sizes. The objective of this work is to investigate the effects of mesh types and grid densities on the hemodynamic properties of physiological flows to determine the most cost-effective meshing approach. The uncertainties of the results will be calculated using the GCI method through steady flow analysis of the aorta models. The grid independence study has been conducted with a constant inlet condition independent of the time step size, focusing on the effects of grid resolution [8, 20]. Additionally, unsteady simulations are performed to determine the appropriate time step based on the mean-TAWSS (Time-Averaged Wall Shear Stress) using the selected element sizes.

## 2. Methodology

### 2.1. Patient-specific geometries

Simulations were conducted on three patient-specific aorta models: a healthy aorta, fusiform aneurysm, and saccular aneurysm. 3D CAD models of the healthy aorta and fusiform aneurysm (Fig. 1a, b) were created using 3D Slicer v5.2, an open-source medical imaging software, from DICOM-formatted CT data obtained from a university hospital [21, 22]. The main step to generate a 3D CAD model in this software is the segmentation process, and it consists of delineating or "segmenting" the areas of interest within the 3D volume. This can be done either manually or semi-automatically through the utilization of different tools available in 3D Slicer. The main objective of segmentation is to isolate anatomical structures or regions from the surrounding tissues. The data was processed using manual segmentation and automatic editing to fill between slices using the software's post-process tool kits. The saccular aneurysm CAD model (Fig. 1c) was also generated using DICOM data from SimVascular's open-source database. Simvascular represents a fully open-source software package that offers a comprehensive pipeline from medical image data segmentation to patient-specific blood flow simulation and analysis [23, 24]. A similar workflow has been followed for the generation of healthy aorta and fusiform aneurysm models as for the generation of saccular aneurysm models.

### 2.2. Discretization method

Three different mesh types, including tetrahedral, polyhedral, and poly-hexacore elements, were used to discretize the aorta models using FLUENT v22 Meshing (ANSYS Inc., Canonsburg, PA). Conventional tetrahedral elements (Fig 2a) and polyhedral elements (Fig. 2b) were used to construct unstructured meshes for all aorta models. The mosaic mesh technology was employed to connect the polyhedral and other mesh types [25]. A poly-hexacore mesh (Fig. 2c), consisting of hexahedral elements in the interior and octahedral elements at the periphery of the models, was generated for grid discretization. Local refinement regions (light gray areas), as shown in Fig. 1, were created to increase accuracy in critical areas, including bifurcations (for all models) and the entrance of the aneurysm (for the saccular model). To resolve high-velocity gradients, the mesh was generated with seven layers in the near-wall region, with a growth rate of 1.2 and a first cell height that resulted in an average y+ value of 1.

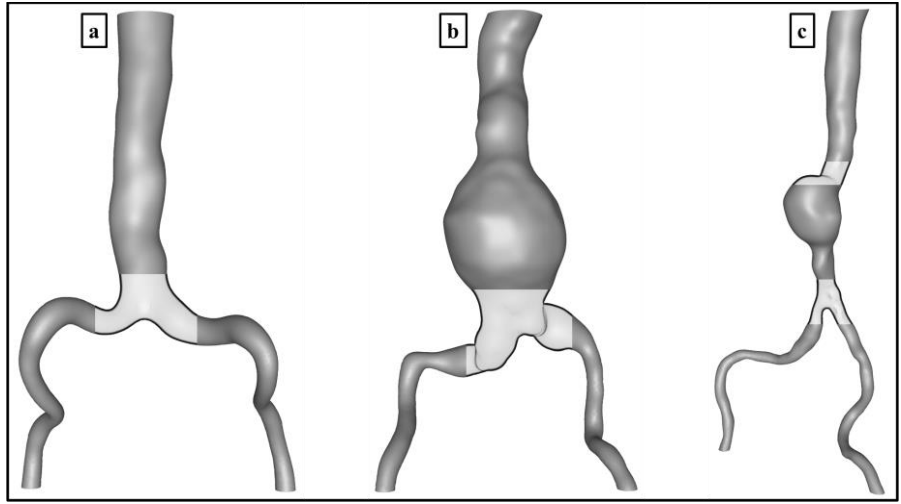

Figure 1: The generated patient-specific geometries with local refinement regions (light gray); (a) healthy aorta, (b) fusiform aneurysm, and (c) saccular aneurysm.

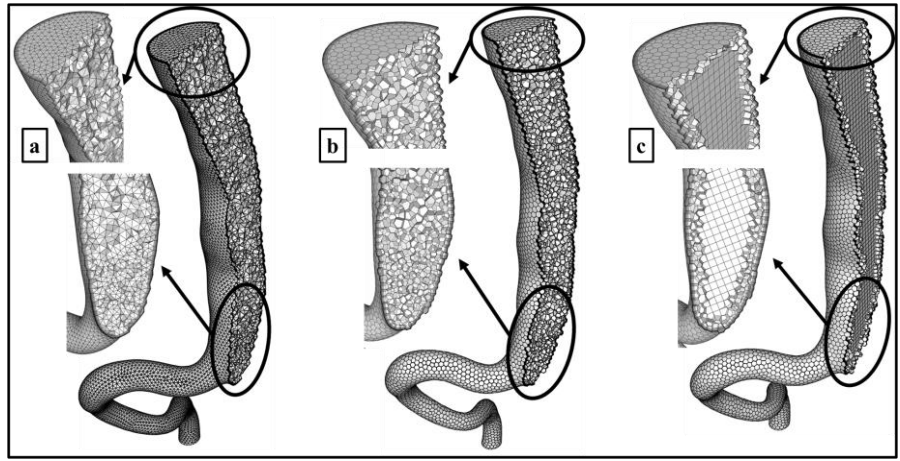

Figure 2: Detailed surface and volumetric mesh with boundary layers for healthy aorta as a sample (a) tetrahedral, (b) polyhedral, and (c) poly-hexacore.

*2.3. Numerical scheme and boundary conditions*

The simulations were performed using ANSYS Fluent v22.1 to numerically solve the RANS equations. The standard k-$\epsilon$ turbulence model was utilized for both steady and unsteady simulations [26]. Steady-state simulations have been performed to determine the suitable cell numbers for each aorta model during mesh independence analyses, while transient simulations have been performed to specify the proper time step sizes. The pressure-velocity coupling was set to the Coupled algorithm for steady simulations and the PISO algorithm for unsteady simulations to solve the momentum and pressure-based continuity equations. The second-order upwind scheme was used to discretize the momentum equations. The normalized continuity and velocity convergence criteria were considered satisfied when the residuals were below $10^{-5}$, and three cardiac cycles were completed for each simulation.

While considering the hemodynamic behavior of blood flow, laminar flow conditions prevail inside the arteries due to the low Reynolds number. However, vortex structures are also formed due to the structure and geometry of blood vessels. In CFD analyses of abdominal aortic aneurysms, both laminar and turbulent models have been used. Vortices can be formed within the aneurysm sac when both models are used [27]. The turbulence model can be employed for transition flows at low Reynolds numbers since the results obtained using the turbulence model are more consistent with experimental results compared to the laminar model [28]. Furthermore, the turbulence model has an advantage over the laminar model, as applying an enhancement wall function yields better performance for flow analyses in the boundary layer close to the artery wall [29].

Table 1: Overview of simulation features and boundary conditions

| Parameters | Conditions / Values |
|---|---|
| *Solver Type* | Pressure-based |
| *Time* | Steady (for mesh independence analyses) |
| | Transient (for time-independence analyses) |
| *Viscous Model* | Standard k-$\epsilon$ |
| *Algorithm for Pressure-Velocity* | Coupled (for steady Simulations) |
| | PISO (for transient Simulations) |
| *Solution Method for Discretization* | Second order |
| *Inlet* | Mass flow rate; |
| | 0.05 *kg/s* (for steady simulations) |
| | Physiological waveform at T=1 *s* (for transient simulations) |
| *Outlet* | P = 0 |
| *Wall* | Rigid, No-slip condition |
| *Operating Pressure* | 100 *mmHg* |
| *Fluid Model* | Blood as a continuous phase |
| *Density of Fluid* | 1060 *kg/m³* |
| *Viscosity of Fluid* | 4 *cP* |

The time-dependent physiological waveform of mass flow rate based on a study was used [30]. For steady flow simulations, a constant mass flow inlet boundary condition (BC) of 0.05 *kg/s*, normal to the boundary direction, was assigned to the inlet of the aorta models. The flow rate waveform was defined using the expression module in Fluent Solver for unsteady analysis. The time-dependent mass flow rate equation was derived using a curve fitting method. Blood was modeled as a homogeneous and incompressible Newtonian fluid. The assumption of a Newtonian fluid for blood is valid for sufficiently large vessels at high shear rates (>100 $s^{-1}$), and the dynamic viscosity and blood density were taken as 4 *cP* and 1060 *kg/m³*, respectively [27, 31].

For all aorta models, the outlet BC was set as a pressure-outlet condition with atmospheric pressure and operating pressure of 100 *mmHg*. The walls were assumed to be rigid with a no-slip condition. The detailed information about the simulation features and defined boundary conditions have been presented in Table 1.

*2.4. Mesh independence analysis*

Different mesh types were assessed using five different grid sizes for the three geometries. The grid sizes were decreased with a refinement factor of 1.3 to increase the mesh density [32, 33]. Table 2 provides detailed information about the cell sizes and mesh numbers.

Table 2: Overview of generated mesh for the aorta models

| | | Cell Sizes | | | Cell Numbers (N) | | | Difference | |
|---|---|---|---|---|---|---|---|---|---|
| | | *Surface* | *Volume (Local) Cell Size [mm]* | *Number of Layers on Near-Wall* | *Tetrahedral* | *Polyhedral* | *Poly-Hexacore* | *Poly. vs. Tet. in (%)* | *Poly-Hexacore vs. Tet.in (%)* |
| **Healthy Aorta** | *Case 5* | 1.18 | 1.78 (1.18) | 7 | 688,730 | 248,862 | 263,864 | -63.9 | -61.7 |
| | *Case 4* | 0.91 | 1.37 (0.91) | 7 | 1,238,446 | 428,683 | 469,396 | -65.4 | -62.1 |
| | *Case 3* | 0.7 | 1.05 (0.7) | 7 | 2,275,508 | 748,859 | 849,004 | -67.1 | -62.7 |
| | *Case 2* | 0.54 | 0.81 (0.54) | 7 | 4,249,263 | 1,346,288 | 1,597,929 | -68.3 | -62.4 |
| | *Case 1* | 0.42 | 0.62 (0.42) | 7 | 8,008,985 | 2,380,600 | 2,915,182 | -70.3 | -63.6 |
| **Fusiform Aneurysm** | *Case 5* | 1.92 | 2.69 (1.92) | 7 | 564,699 | 188,640 | 199,067 | -66.6 | -64.7 |
| | *Case 4* | 1.48 | 2.07 (1.48) | 7 | 1,048,462 | 333,347 | 362,997 | -68.2 | -65.4 |
| | *Case 3* | 1.14 | 1.59 (1.14) | 7 | 1,994,107 | 597,662 | 669,451 | -70 | -66.4 |
| | *Case 2* | 0.88 | 1.23 (0.88) | 7 | 3,814,372 | 1,077,820 | 1,243,932 | -71.7 | -67.4 |
| | *Case 1* | 0.68 | 0.95 (0.68) | 7 | 7,462,052 | 2,004,312 | 2,391,147 | -73.1 | -68 |
| **Saccular Aneurysm** | *Case 5* | 1.18 | 1.78 (1.18) | 7 | 649,409 | 240,039 | 255,800 | -63 | -60.6 |
| | *Case 4* | 0.91 | 1.37 (0.91) | 7 | 1,140,063 | 403,925 | 446,397 | -64.6 | -60.8 |
| | *Case 3* | 0.7 | 1.05 (0.7) | 7 | 2,087,390 | 716,260 | 827,629 | -65.7 | -60.4 |
| | *Case 2* | 0.54 | 0.81 (0.54) | 7 | 3,839,624 | 1,257,267 | 1,514,570 | -67.3 | -60.6 |
| | *Case 1* | 0.42 | 0.62 (0.42) | 7 | 7,113,107 | 2,213,386 | 2,788,122 | -68.9 | -60.8 |

### 2.4.1. Grid convergence index method

In this study, numerical uncertainties resulting from spatial discretization and iterative convergence errors were quantified using the Grid Convergence Index method [13]. This method is based on generalized Richardson extrapolation [34] and compares the results of at least two different grid sizes. Richardson extrapolation was applied as proposed by Celik and Zhang [35] to the simple turbulent flows. Celik and Karatekin [11] have also reported using the same method for nonuniform grids. The error based on the Richardson extrapolation stated as

$$E = f(h) - f_{exact} = Ch^p + H.O.T. \tag{1}$$

where $E$ is the discretization error, $f(h)$ is the grid solution at two different discrete spacing of $h$. Its analytical solution is $f_{exact}$, $C$ is a constant coefficient, $p$ is the order of convergence, and $H.O.T.$ is the higher-order terms. The order of convergence can obtain with

$$p = \left| \ln \left( \frac{f_k - f_j}{f_j - f_i} \right) \right| / \ln(r) \tag{2}$$

where

$$r = \frac{h_k}{h_j} = \frac{h_j}{h_i} \tag{3}$$

where $r$ is the grid refinement ratio. $i$, $j$, and $k$ represent fine, medium, and coarse grid resolutions, respectively. The discretization error is used to calculate an error band, and this is expressed for the fine grid solution as

$$GCI_{ij} = F_s \cdot E_{ij} \tag{4}$$

where $F_s$ is a safety factor. Its value was set by Roache [36] has as 3.0 for two different grid spacings, and 1.25 for three or more. In Eq. (4), is computed as

$$E_{ij} = \frac{e_{ij}}{r^p - 1} \tag{5}$$

and relative error,

$$e_{ij} = \left| \frac{f_j - f_i}{f_i} \right| \tag{6}$$

where $f_i$ and $f_j$ are the grid solutions for the fine and medium grid resolutions, respectively. The continuum value at the zero-grid spacing $f_{h=0}$ can replace the analytical solution $f_{exact}$ when it is unknown in the calculation of the actual fractional error $e_0$ [37]:

$$e_0 = \frac{f(h) - f_{h=0}}{f_{h=0}} \tag{7}$$

where

$$f_{h=0} \quad f_i + \frac{f_i - f_j}{r^p - 1} \tag{8}$$

The following equality is used to check whether the numerical convergence is within the asymptotic range to ensure mesh independence

$$\frac{GCI_{jk}}{r^p GCI_{ij}} \approx 1 \tag{9}$$

$$R^* = \frac{f_j - f_i}{f_k - f_j} \tag{10}$$

Finally, note that the apparent convergence condition must be within the monotonic convergence range to use the generalized Richardson extrapolation method. It is calculated using Eq. (10) and $R^*$ value of the convergence condition is determined as follows:

$R^* > 1$         Monotonic divergence
$1 > R^* > 0$    Monotonic convergence
$0 > R^* > -1$   Oscillatory convergence
$R^* < -1$        Oscillatory divergence

Table 3: Mesh independence analysis using the GCI method based on wall shear stress

| | Mesh Type | Cell Numbers (N) | | | Cases (i j k) | GCI_ij (%) | GCI_jk (%) | Asymptotic Range of Convergence |
|---|---|---|---|---|---|---|---|---|
| **Healthy Aorta** | Tetrahedral | 8,008,985 | 4,249,263 | 2,275,508 | 1 2 3 | 0.0043 | 0.0009 | 1.0022 |
| | | 4,249,263 | 2,275,508 | 1,238,446 | 2 3 4 | 0 | 0 | 1.0005 |
| | | 2,275,508 | 1,238,446 | 688,730 | 3 4 5 | 0.1085 | 0.0805 | 1.006 |
| | Polyhedral | 2,380,600 | 1,346,288 | 748,859 | 1 2 3 | 0.0002 | 0.0005 | 0.9993 |
| | | 1,346,288 | 748,859 | 428,683 | 2 3 4 | 0.0066 | 0.011 | 0.9977 |
| | | 748,859 | 428,683 | 248,862 | 3 4 5 | 0.0059 | 0.0111 | 0.9962 |
| | Poly-Hexacore | 2,915,182 | 1,597,929 | 849,004 | 1 2 3 | 0.0008 | 0.0018 | 0.999 |
| | | 1,597,929 | 849,004 | 469,396 | 2 3 4 | 0.0099 | 0.0152 | 0.9977 |
| | | 849,004 | 469,396 | 263,864 | 3 4 5 | 0.0043 | 0.0086 | 0.9965 |
| **Fusiform Aneurysm** | Tetrahedral | 7,462,052 | 3,814,372 | 1,994,107 | 1 2 3 | 0.002 | 0.0042 | 1.0019 |
| | | 3,814,372 | 1,994,107 | 1,048,462 | 2 3 4 | 24.3191 | 24.7649 | 1.004 |
| | | 1,994,107 | 1,048,462 | 564,699 | 3 4 5 | 0.6522 | 0.7125 | 1.0041 |
| | Polyhedral | 2,004,312 | 1,077,820 | 597,662 | 1 2 3 | 0.0022 | 0.004 | 0.9988 |
| | | 1,077,820 | 597,662 | 333,347 | 2 3 4 | 0.0028 | 0.0057 | 0.9977 |
| | | 597,662 | 333,347 | 188,640 | 3 4 5 | 0.0118 | 0.02 | 0.9954 |
| | Poly-Hexacore | 2,391,147 | 1,243,932 | 669,451 | 1 2 3 | 0.0067 | 0.0112 | 0.9976 |
| | | 1,243,932 | 669,451 | 362,997 | 2 3 4 | 14.5842 | 14.2573 | 0.996 |
| | | 669,451 | 362,997 | 199,067 | 3 4 5 | 0.0177 | 0.0269 | 0.9961 |
| **Saccular Aneurysm** | Tetrahedral | 7,113,107 | 3,839,624 | 2,087,390 | 1 2 3 | 0.0069 | 0.0106 | 1.0017 |
| | | 3,839,624 | 2,087,390 | 1,140,063 | 2 3 4 | 0.0858 | 0.0694 | 1.0026 |
| | | 2,087,390 | 1,140,063 | 649,409 | 3 4 5 | 0.0019 | 0.0041 | 1.0021 |
| | Polyhedral | 2,213,386 | 1,257,267 | 716,260 | 1 2 3 | 0.0022 | 0.0044 | 0.9981 |
| | | 1,257,267 | 716,260 | 403,925 | 2 3 4 | 0.0057 | 0.0109 | 0.9962 |
| | | 716,260 | 403,925 | 240,039 | 3 4 5 | 0.6008 | 0.6696 | 0.9928 |
| | Poly-Hexacore | 2,788,122 | 1,514,570 | 827,629 | 1 2 3 | 0.0025 | 0.0049 | 0.9982 |
| | | 1,514,570 | 827,629 | 446,397 | 2 3 4 | 0.005 | 0.0096 | 0.9964 |
| | | 827,629 | 446,397 | 255,800 | 3 4 5 | 0.4116 | 0.4682 | 0.9931 |

*2.4.2. Time independence analysis*

After the most suitable mesh type and mesh size is determined using the *GCI* method, time independence analysis was performed to check the effects of the time step on the results by evaluating the mean time-averaged wall shear stress (*TAWSS*). Four different time steps (1, 2, 5, and 10 *ms*) have been specified to conduct unsteady simulations with selected appropriate mesh sizes. The time step sizes have been determined by considering the Courant Number, and it is generally suggested to keep the Courant number below 1 to ensure stability and accuracy in the numerical solution. However, the range of Courant Numbers in this study was designated between 0.07 and 2.97 for each time step size of all geometries in order to strike a trade-off between accuracy and computational efficiency.

## 3. Results

*3.1. Computational performance*

The analysis of aortic flow using Poly and Poly-Hexacore mesh types required significantly less iterations when compared to the Tet mesh type. The primary reason for this difference is the variation in mesh density between the grid types. Specifically, Poly and Poly-Hexacore mesh types lead approximately 60 to 70% fewer cell numbers than the Tet type when considering the same grid sizes for all geometries. Consequently, the number of iterations required for the Tet mesh type are significantly higher compared to the other meshes and ranging between 85 to 275%.

As anticipated, the number of cells directly impacts the total CPU time required for the simulations. The Tet mesh type exhibits a significantly higher computation cost, ranging from 100 to 550% higher than for both Poly and Poly-Hexacore and this is for all aorta geometries. Furthermore, the Poly grid type is computationally more efficient compared to Poly-Hexacore.

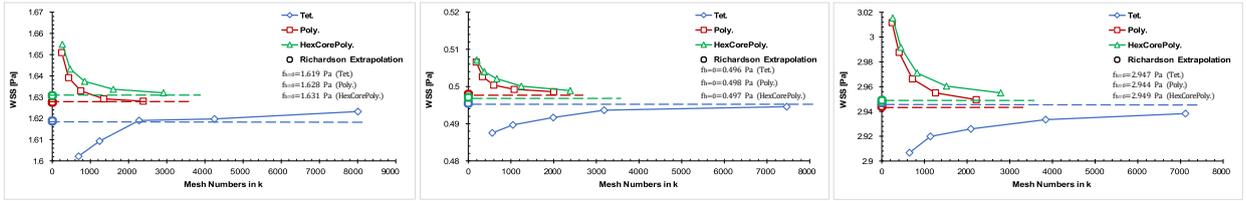

Figure 3: Wall Shear Stress (WSS) for different mesh types and sizes with Richardson Extrapolation (a) healthy aorta, (b) fusiform aneurysm, and (c) saccular aneurysm.

*3.2. GCI analysis*

The GCI analysis evaluated the successive grid refinement and quality for each mesh type, focusing on the GCI values based on wall shear stress. Table 3 shows different convergence conditions influenced by the uncertainties associated with the discretization method and systematic grid refinement. The grid sizes were consistently refined from coarse to fine, with an r-value of 1.3 ($r_{ij} = r_{jk}$).

Compared to the other mesh types, the Tet mesh type exhibited non-monotonic convergence or varying divergence conditions for the wall shear stress. Notably, an extreme order of convergence value, such as $p=18.77$ for the medium mesh triplet of Tet, was observed alongside a small GCI value. This observation raises concerns about the reliability of the uncertainty range. According to GCI analyses, only the monotonic condition for convergence and divergence has been observed, and no oscillatory condition has been reached. In general, the results have reached monotonic convergence for all geometries and mesh types. However, the analyses for tetrahedral mesh types have shown a monotonic divergence situation at a few points, and these points have been discarded during the assessment. Additionally, there is a strong relationship between mesh refinement and the convergence condition. As cell sizes decrease through refinement, the results have more easily reached the monotonic convergence condition, especially for polyhedral mesh types.

For the healthy aorta and the saccular aneurysm, all Poly and Poly-Hexacore mesh triplets achieved the asymptotic range with monotonic convergence conditions. However, the Tet mesh style demonstrated divergence conditions despite reaching the asymptotic range of convergence for certain mesh triplets.

In the case of the fusiform aneurysm, the Poly mesh type displayed only monotonic convergence conditions for each evaluation value, while Tet and Poly-Hexacore exhibited various convergence and divergence conditions. Furthermore, the Poly mesh type reached the asymptotic range in the finest two triplets. Additionally, as another notable result of the analysis the grid convergence evaluation of this geometry reveals that the medium mesh triplet of Tet and Poly-Hexacore exhibit high GCI values of 24% and 14% for WSS, respectively.

Wall shear stress is an important parameter associated with AAA as it provides valuable insight into the risk of aneurysm formation, growth, and rupture. The distribution of WSS along the aorta wall has been compared using different discretization methods, namely Tet, Poly, and Poly-Hexacore. The WSS results have been obtained by refining successive grids for each mesh style, as depicted in Fig. 3.

The trend of the findings has been analyzed based on the estimation of analytical results. The analytical results were predicted using Eq. (8), utilizing the solutions of the two finest grids. Richardson extrapolation was performed to estimate the continuum value of $f_{h=0}$ at zero grid spacing [13, 33]. The continuum values obtained for each mesh type were very close, with slight differences.

The actual WSS values approach the estimated exact results for the healthy aorta, fusiform aneurysm, and saccular aneurysm, with values of approximately 1.62 *Pa*, 0.49 *Pa*, and 2.94 *Pa*, respectively. The trend of WSS values for Poly and Poly-Hexacore shows a decreasing trend as the grid sizes become smaller, while Tet exhibits an increasing trend. The normalized numerical results, obtained using the values at zero grid spacing of WSS, are shown in Fig. 4, indicating the asymptotic regions.

For the Tet mesh type, the normalized values for the geometries range from 0.98 to 1.0 from coarse to the finest mesh, while for Poly and Poly-Hexacore, the values decrease from 1.02 to approximately 1.0. The actual fractional error of the grid solutions, as shown in Fig. 5, has been calculated using Eq. (6) as an assessment criterion. The percentage errors range from -2% to 2.5% for all models.

*3.3. Wall shear stress*

The distribution of WSS along the aorta wall has been compared using different discretization methods, namely Tet, Poly, and Poly-Hexacore. The WSS results have been obtained by refining successive grids for each mesh style, as depicted in Fig. 3.

The trend of the findings has been analyzed based on the estimation of analytical results. The analytical results were predicted using Eq. (8), utilizing the solutions of the two finest grids. Richardson extrapolation was performed to estimate the continuum value of $f_{h=0}$ at zero grid spacing [13, 33]. The continuum values obtained for each mesh type were very close, with slight differences.

The actual WSS values approach the estimated exact results for the healthy aorta, fusiform aneurysm, and saccular aneurysm, with values of approximately 1.62 *Pa*, 0.49 *Pa*, and 2.94 *Pa*, respectively. The trend of WSS values for Poly and Poly-Hexacore shows a decreasing trend as the grid sizes become smaller, while Tet exhibits an increasing trend. The normalized numerical results, obtained using the values at zero grid spacing of WSS, are shown in Fig. 4, indicating the asymptotic regions.

For the Tet mesh type, the normalized values for the geometries range from 0.98 to 1.0 from coarse to the finest mesh, while for Poly and Poly-Hexacore, the values decrease from 1.02 to approximately 1.0. The actual fractional error of the grid solutions, as shown in Fig. 5, has been calculated using Eq. (6) as an assessment criterion. The percentage errors range from -2% to 2.5% for all models.

*3.4. Time independence*

The assumed heart rate for transient analyses, as indicated in Table 1, is 60 beats per minute, with each cycle lasting one second. Simulations were conducted over three cardiac cycles. For time-independent analyses, the simulation outcomes were evaluated after two cardiac cycles to mitigate initial effects. Time-Averaged Wall Shear Stresses (TAWSS) were computed throughout the duration of the final cardiac cycle (third cycle) by using various time step (TS) sizes. TS sizes of 1, 2, 5, and 10 *ms* were selected for the unsteady simulations. The results, obtained from the third cardiac cycle, are summarized in Table 4. All simulations with different TS sizes have achieved iteration convergence. The differences in mean TAWSS results between the simulations are minimal and negligible. However, increasing the TS size leads to a significant rise in computational time. Therefore, selecting the shortest and most appropriate time step size for unsteady simulations is crucial, and an optimal time step of 10 *ms* has been chosen (grey columns in Table 4).

Table 4: Summary of the mean-TAWSS results for selected grid types and sizes

| | Healthy Aorta | | | | Fusiform Aneurysm | | | | Saccular Aneurysm | | | |
|---|---|---|---|---|---|---|---|---|---|---|---|---|
| | $\Delta t_1$ | $\Delta t_2$ | $\Delta t_3$ | $\Delta t_4$ | $\Delta t_1$ | $\Delta t_2$ | $\Delta t_3$ | $\Delta t_4$ | $\Delta t_1$ | $\Delta t_2$ | $\Delta t_3$ | $\Delta t_4$ |
| **Time Step Sizes** | 1 ms | 2 ms | 5 ms | 10 ms | 1 ms | 2 ms | 5 ms | 10 ms | 1 ms | 2 ms | 5 ms | 10 ms |
| **Mean-TAWSS** | 1.74096 | 1.74020 | 1.73797 | 1.73444 | 0.54004 | 0.53984 | 0.53928 | 0.53836 | 3.15563 | 3.15401 | 3.14925 | 3.14171 |
| **Difference** | - | -0.04% | -0.13% | -0.20% | - | -0.04% | -0.10% | -0.17% | - | -0.05% | -0.15% | -0.24% |
| **Selected Mesh Type** | Polyhedral | | | | Polyhedral | | | | Polyhedral | | | |
| **Selected Number of Elements** | 748,859 (Case 3) | | | | 1,077,820 (Case 2) | | | | 716,260 (Case 3) | | | |

## 4. Discussion

Grid resolution plays a crucial role in investigating the hemodynamics of aorta models and assessing the uncertainties associated with each mesh type. The reliability of WSS computations depends on achieving convergence and being within the asymptotic range. The gray cells in Table 3 indicate that the solutions have reached the asymptotic regions, which have been defined based on the asymptotic range of convergence and monotonic convergence conditions. A percentage difference of 2% has been used as a criterion for commenting on the asymptotic regions. It is essential for accuracy that the results fall within the asymptotic range and exhibit monotonic convergence [8, 9]. The findings have indicated that the accuracy of the GCI values is significantly influenced by the order of convergence [38, 39].

The continuum value at zero grid spacing can be estimated using Eq. (8) (see WSS in Fig. 3) with $h^2$ extrapolation [34] when the exact analytical solution is unknown. This estimation allows for the normalization of grid solutions, as shown in Fig. 4. While the normalized grid solutions for all Tet mesh type models are around 1, they have not reached the asymptotic range based on the WSS results due to convergence conditions. Poly and Poly-Hexacore exhibit similar tendencies in the normalized results when successive grid refinements are considered. The WSS values approach 1 with all grids.

Similarly, the actual fractional errors of the Tet mesh type for all aorta models are within a difference of 2% for WSS. However, only the finest triplet mesh has reached the asymptotic range for fusiform and saccular aneurysms. The findings and their fractional errors for Poly and Poly-Hexacore have a difference of 2% in terms of WSS when all models are considered.

Transient simulations were performed for time independence analysis, using a physiological waveform of flow rate at the inlet boundary condition instead of a constant value. The results demonstrate distinctive differences in wall shear stress between steady and pulsatile flow analyses [8]. The mean-TAWSS data from the transient simulations, performed using various time-step sizes, provides more accurate results than the data from steady-state simulations.

## 5. Conclusion

This study aimed to assess the effectiveness of different discretization techniques and their impact on the hemodynamics of diseased (fusiform and saccular aneurysm) and healthy aorta. Additionally, it introduced the first mesh independence analysis of aortic flow simulations based on GCI method to evaluate uncertainties associated with discretization, iterations, and solution methods for patient-specific simulations.

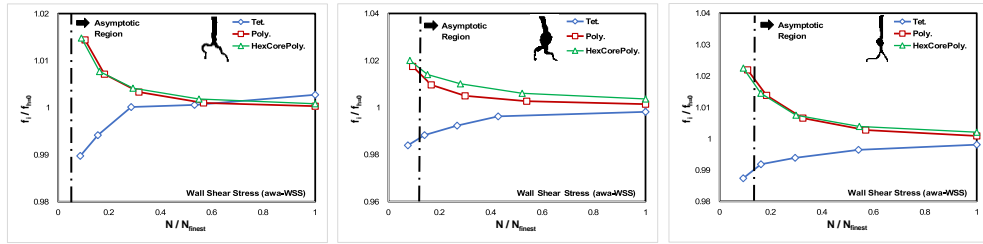

Figure 4: The normalized grid solutions using the continuum values for WSS.

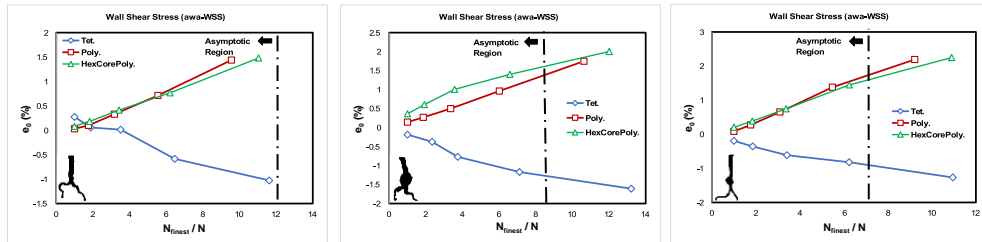

Figure 5: Actual fractional error with normalized grid sizes for WSS.

The results clearly demonstrated that appropriate grid resolution significantly influences the accuracy and reliability of WSS results. Based on the findings and GCI analysis, the polyhedral mesh type was selected for all patient-specific aorta models. It was clear that it is more suitable than other mesh types when considering uncertainties and computational costs associated with the mesh styles.

**Limitations**

Several limitations need to be acknowledged in this study. The simulations differ from the *in − vivo* conditions due to certain assumptions and restrictions. The aorta walls were assumed to be rigid, and the outlet boundary conditions were set as $P_{outlet} = 0$ under operating pressure $P_{operating} = 100$ *mmHg*. These limitations should be addressed in future research to validate the findings against experimental measurements.

**Acknowledgments**